\documentclass[twocolumn]{openjournal}

\usepackage{xcolor}
\usepackage{textgreek}
\usepackage[utf8]{inputenc}
\usepackage[english]{babel}
\usepackage{hyperref}
\usepackage{dcolumn}
\usepackage{bm}
\usepackage[caption=false]{subfig}
\usepackage{placeins}
\usepackage{tikz}
\usepackage{amsmath}
\usepackage{float}
\usepackage{adjustbox}
\usepackage{enumitem}

\usetikzlibrary{shapes.geometric, arrows}
\tikzstyle{startstop} = [rectangle, rounded corners, minimum width=3cm, minimum height=1cm,text centered, draw=black, fill=red!30]
\tikzstyle{process} = [rectangle, minimum width=3cm, minimum height=1cm, text centered, draw=black, fill=blue!30]
\tikzstyle{arrow} = [thick,->,>=stealth]
\tikzstyle{decision} = [diamond, aspect=3, minimum width=2.5cm, minimum height=1cm, text centered, draw=black, fill=green!30]
\definecolor{keywordcolor}{rgb}{0.0, 0.0, 1.0}
\definecolor{commentcolor}{rgb}{0.5, 0.5, 0.5}
\definecolor{stringcolor}{rgb}{0.0, 0.5, 0.0}
\hypersetup{
    colorlinks=true, % Enable colored links
    citecolor=blue,   % Color of citation links (e.g., bibliography)
    urlcolor=magenta % Color of external links (e.g., URLs)
}

\newcommand{\etal}{\textit{et al}.}

\begin{document}

\title{Detecting wide binaries using machine learning algorithms}

\author{Amoy Ashesh}
\email{ashesha@tcd.ie}
\affiliation{Department of Physics, Indian Institute of Technology Patna, Patna, Bihar 801106, India}
\affiliation{Department of Physics, Trinity College Dublin, The University of Dublin, Dublin 2, Ireland}

\author{Harsimran Kaur}
\email{05harsimran@gmail.com}
\affiliation{Department of Physics, Indian Institute of Technology Patna, Patna, Bihar 801106, India}

\author{Sandeep Aashish}
\email{aashish@iitp.ac.in}
\affiliation{Department of Physics, Indian Institute of Technology Patna, Patna, Bihar 801106, India}

\begin{abstract}
We present a machine learning (ML) framework for the detection of wide binary star systems using Gaia DR3 data. By training supervised ML models on established wide binary catalogues, we efficiently classify wide binaries and employ clustering and nearest neighbour search to pair candidate systems. Our approach incorporates data preprocessing techniques such as SMOTE, correlation analysis, and PCA, and achieves high accuracy and recall in the task of wide binary classification. The resulting publicly available code enables rapid, scalable, and customizable analysis of wide binaries, complementing conventional analyses and providing a valuable resource for future astrophysical studies.
\end{abstract}

% \keywords{Computational Astrophysics, Machine Learning, Statistical Learning, Clustering, Wide-Binary Systems, Anomaly Detection}

\maketitle

%\noindent \textbf{Keywords:} Computational Astrophysics, Machine Learning, Statistical Learning, Clustering, Wide-Binary Systems, Anomaly Detection.

\section{Introduction}

Machine Learning (ML) has evolved into one of the most pivotal tools in the era of data intensive astronomy due to its efficiency and scalability, and is set to play a key role in the search for new physics in the coming decades. In recent literature, various studies have employed machine learning techniques to extract information from raw data which is otherwise difficult to analyse analytically and often computationally expensive. Stellar classification on the SIMBAD database was studied in Ref. \cite{Cody_2024}, classification of accretion states of black holes was studied in Ref. \cite{Sreehari_2021}, imposing constraints on the deviations from general relativity using ML in Ref. \cite{Alestas_2022}, the detection and parameter estimation process of gravitational waves using ML was carried out in Ref. \cite{Koloniari_2025}. For an exhaustive summary of recent works in this directions, see Refs. \cite{baron2019,haghighi2023,li2025}. 

In this paper, we take the first steps to introduce machine learning assisted search for new physics in the recently released Gaia DR3 dataset through the classification of wide binaries. The wide binary classification problem is among the well-known classification problems in astronomy, and extensively studied in literature using both traditional statistical methods \cite{El_Badry_2021,banik2018,chae2023robustevidencebreakdownstandard} and machine learning techniques \cite{Sreehari_2021,Cody_2024,li2025} in different contexts. The classification problem of wide binaries is interesting because these are gravitationally bound pairs of stars with large separations and can be used to study stellar evolution, dynamics, galactic structure, as well as potential signatures of deviations from standard gravity \cite{Hernandez_2012}. Wide binary pairs of stars separated by thousands to tens of thousands of astronomical units operate precisely in the low-acceleration regime where modified gravity effects might emerge. Recent Gaia data releases have provided an unprecedented opportunity to study these systems across the Galaxy with high precision. However, identifying true gravitationally bound pairs and detecting subtle anomalies in their dynamics is complicated by noise, contamination and the scale of the dataset, thereby necessitating complex statistical analysis \cite{El_Badry_2021}.

We have employed a supervised ML approach to predict wide binaries. As is standard in any machine learning framework, various data preprocessing techniques like correlation analysis, Synthetic Minority Oversampling Technique (SMOTE) and Principal Component Analysis (PCA) have been employed in this work. Confusion matrices and standard ML metrics have been used to analyse the performance of the models and tune the hyperparameters accordingly. The codes used in this work are made available as a set of publicly available tools (hosted at \href{https://github.com/DespCAP/G-ML}{https://github.com/DespCAP/G-ML}) which can be used to generate a catalogue of wide binaries using our pre-trained models, or to train the models locally.

The structure of this paper is as follows. Sec. \ref{mlte} outlines the essential machine learning tools and techniques, including those of data preprocessing and evaluation of models. Sec. \ref{wbcl} describes the methodologies for using ML algorithms to predict wide binary pairs from the Gaia DR3 dataset. We conclude with a few remarks and a future outlook in Sec. \ref{conclusion}.

\section{Machine Learning Techniques}
\label{mlte}
The models are fitted onto a training dataset. The dataset utilised for prediction is called the testing dataset. The dataset was split into train and test cases in the ratio of 80:20 (train:test). The various Machine Learning models used are:
\begin{enumerate}
\item Logistic Regression:
In machine learning, the Supervised Learning subcategory includes the commonly used algorithm of logistic regression. Its primary purpose is to predict the outcome of a dependent variable that belongs to a category based on a set of independent variables. This implies that the output must be categorical or discontinuous, such as Yes or No, 0 or 1, or true or false \cite{cox2018}. Nevertheless, rather than offering a precise value of 0 or 1, logistic regression generates probability values that fall within the range of 0 to 1.
\item Decision Tree Classifier:
A supervised ML algorithm that is utilized primarily for classification tasks, although it can also solve regression problems. It operates on a tree-like structure that includes internal nodes representing the characteristics of a given data-set \cite{ROKACH2016}. The branches denote the decision-making processes, and the leaf-nodes indicate the result. A DT comprises of two types of nodes: Decision Nodes, that possess multiple branches and are responsible for decision-making, and Leaf Nodes, which lack branches and represent the final decision or output. The Decision Tree arrives at its decisions or tests based on the properties or characteristics of the provided dataset.\cite{ROKACH2016}
\item Random Forest Classifier:
A supervised ML algorithm, helps classify the output variable as categorical or discontinuous. An RFC is based on ensemble learning, combining multiple decision trees to make more accurate predictions.\cite{Breiman2001} The algorithm creates a forest of DTs, each using a random subset (RSS). Each RSS has different features and data points.
During the training process, the RFC randomly selects a subset of features, and the RSS creates a decision tree. This process is repeated several times to create multiple decision trees. \cite{Breiman2001} The algorithm predicts by aggregating each tree's predictions and choosing the class that receives the most votes. This approach helps to improve the accuracy and robustness of the model, as it reduces the impact of individual trees that may be overfitting the data. An RFC is often used for the classification of images or text.
\item K-Nearest Neighbors:
An ML algorithm used for the purposes of classification and regression. It falls under the category of supervised learning, meaning that it requires labelled data to train the model. It identifies new data-points on the basis of proximity to the k-nearest data-points in the training dataset \cite{cover1967}. The user determines the value of ‘k’ and ascertains the number of neighbours to consider. KNN is a non-parametric algorithm, meaning it makes no assumptions about the data distribution \cite{cover1967}. It is also easy to understand and implement, making it a popular choice for many classification and regression tasks. However, its performance can be affected by the choice of k, and it can be computationally expensive for large datasets.
\item Support Vector Machine:
An ML algorithm, helps in tasks related to classification, regression, as well as outlier detection. It is a supervised learning algorithm; it thus requires labelled data to train the model. In SVM, the algorithm constructs a hyperplane (HP) in a High Dimensional Space (HDS) that may be deployed to separate the different classes in the data. \cite{Cortes1995} The objective is to ascertain the HP that maximises the margin, which is defined as the distance between the HP and the nearest data points of each class. 
Using the kernel trick (K-T) technique, SVM can handle both non-linearly and otherwise (linear) separable data (N-/LSD). The K-T transforms the input data into an HDS that can be LSD. SVM is particularly useful when dealing with high-dimensional data, for example, classifying images or text. It helps in handling datasets with a small number of samples, as it is less prone to overfitting compared to other algorithms. However, SVM may not be advisable for larger datasets and can be subject to the specific kernel function as well as other hyperparameters.\cite{Cortes1995} Nonetheless, with careful tuning of the parameters, SVM can be a powerful tool for solving many classification, regression, and outlier detection problems.
\end{enumerate}

\subsection{Evaluation Metrics}
The accuracy, recall and F1 measure are evaluated on each ML algorithm. Confusion Matrices for all the algorithms were also plotted. Accuracy is a metric used to determine the frequency with which a model accurately predicts the outcome of a given task. It is indicated as the ratio of the correct predictions versus the overall predictions \cite{SOKOLOVA2009}. It is particularly useful when the classes in the data are evenly distributed. Recall is a measure of how well the model identifies positive instances. To compute this metric, the sum of true positives is divided by that of true positives and false negatives \cite{SOKOLOVA2009}. It is a valuable metric for correctly identifying all positive instances, such as in medical diagnosis. F1 measure is a combination of precision and recall, which provides a balance between these two metrics. This metric is determined by calculating the harmonic mean of precision and recall \cite{SOKOLOVA2009}. It considers false positives as well as false negatives. F1 measure is often deployed in binary classification problems when the data is imbalanced.

A confusion matrix is a tabular representation that is utilized to assess the effectiveness of a classification model (CM). It is a matrix that summarises the predicted and actual classifications of a model’s output, providing a more detailed view of its performance than just a single accuracy score. A confusion matrix comprises four primary components that are utilised to assess the performance of a classification model. These components are true positives (TPs), false positives (FPs), true negatives (TNs), and false negatives (FNs). Each of them carries a specific meaning. TPs signify the number of instances where the CM accurately predicts the positive-class. FPs are the number of instances where the CM predicts the positive-class despite the actual class being negative. TNs signify the number of instances where the CM accurately predicts the negative-class. FNs correspond to the number of instances where the CM predicts the negative-class despite the actual class being positive \cite{FAWCETT2006}.

\subsection{Data Preprocessing}
In data analysis and machine learning, having an imbalanced dataset can significantly impact the accuracy of the resulting predictions. In such cases, Synthetic Minority Oversampling Technique (SMOTE) is a commonly used method to balance the dataset. SMOTE is a technique that generates synthetic data points for the minority class to balance the distribution of the classes in the dataset. This technique creates new observations for the minority class by using interpolation methods to create "synthetic" samples that are similar to the existing minority class observations \cite{Chawla_2002}. This process continues until the minority class has a representation similar to that of the majority class. By using SMOTE to balance the dataset, the resulting distribution of the classes is more even, which allows for more accurate predictions by machine learning models. This process can mitigate the issue of imbalanced classes and can lead to better results when working with imbalanced datasets. SMOTE, thereby, helps improve the accuracy of ML models when working with imbalanced datasets, and it is frequently used in data analysis and machine learning projects.

\begin{figure}[h!]
    \centering
    \subfloat[Before SMOTE\label{fig:before_smote}]{
        \includegraphics[width=0.5\linewidth]{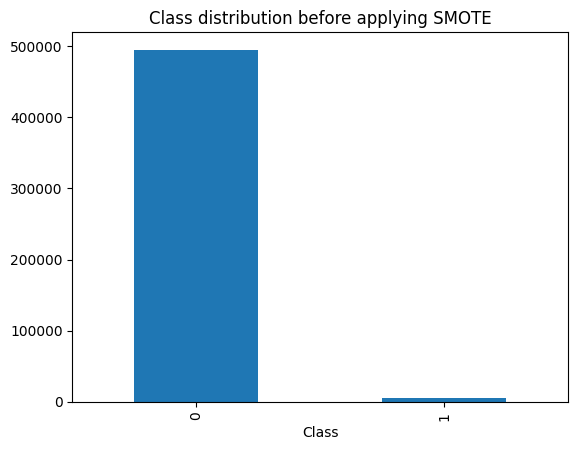}
    }
    \hfill
    \subfloat[After SMOTE\label{fig:after_smote}]{
        \includegraphics[width=0.5\linewidth]{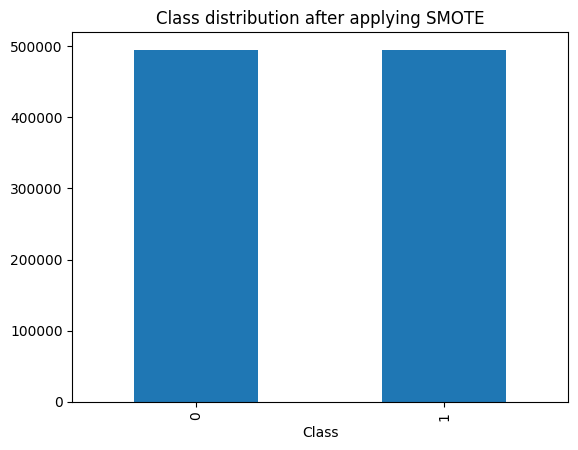}
    }
    \caption{Comparison of data distribution before and after applying SMOTE. Here 0 depicts that the entity is not a part of a WBS and 1 depicts that the entity is a part of a WBS.}
    \label{fig:smote_comparison}
\end{figure}

% \begin{figure}[h!]
%     \centering
%     % First image
%     \begin{subfigure}{0.4\linewidth}
%         \centering
%         \includegraphics[width=\linewidth]{before_SMOTE.png}
%         \caption{Before SMOTE}
%         \label{fig:before_smote}
%     \end{subfigure}
%     \hfill
%     % Second image
%     \begin{subfigure}{0.4\linewidth}
%         \centering
%         \includegraphics[width=\linewidth]{after_SMOTE.png}
%         \caption{After SMOTE}
%         \label{fig:after_smote}
%     \end{subfigure}
%     % Common caption
%     \caption{Comparison of data distribution before and after applying SMOTE. Here 0 depicts that the entity is not a part of a WBS and 1 depicts that the entity is a part of a WBS.}
%     \label{fig:smote_comparison}
% \end{figure}
\FloatBarrier

All the ML models were trained on the SMOTE-balanced as well as the raw-filtered dataset. The ML models, once trained on the SMOTE-balanced dataset, were tested on the SMOTE-balanced test dataset and the raw-filtered test dataset. As depicted in FIG. \ref{fig:smote_comparison}, there was a significant increase in accuracy and other performance metrics corresponding to each ML model. The reason for the marked increase is that the raw-filtered dataset contains very sparse entries, and the desired classes have a stark distinction. Therefore, during the training process, the  ML models inherently develop a bias towards the class with a higher occurring frequency and introduce redundancies that have to be countered by training the ML models on the class-balanced dataset using the SMOTE technique \cite{Chawla_2002}. 

Correlation analysis has been employed to quantify the degree of linear association between two continuous variables. The Pearson correlation coefficient \cite{Benesty2009}, denoted as \( r \), has commonly been used for this purpose. It is defined as:

\begin{equation}
r = \frac{\sum_{i=1}^{n} (x_i - \bar{x})(y_i - \bar{y})}{\sqrt{\sum_{i=1}^{n} (x_i - \bar{x})^2} \sqrt{\sum_{i=1}^{n} (y_i - \bar{y})^2}},
\end{equation}

where \( x_i \) and \( y_i \) represent individual data points, and \( \bar{x} \) and \( \bar{y} \) denote their respective means. The coefficient \( r \) ranges from \( -1 \) to 0 to \( 1 \), indicating perfect negative correlation, no correlation, and perfect positive correlation respectively.

Prior to computing the correlation coefficient, data sets were inspected for normality and linearity, as the Pearson metric assumes both. In cases where these assumptions were violated, the Spearman rank correlation coefficient \cite{de_Winter_2016}, a non-parametric alternative, was used instead. This approach relies on ranked data and measures monotonic relationships, regardless of linearity.

Significance of the correlation has been assessed through hypothesis testing \cite{Rainio2024}, with the null hypothesis assuming no correlation between the variables (\( r = 0 \)). A two-tailed p-value has been calculated to determine whether the observed correlation differs significantly from zero, given the sample size.

Interpretation of correlation results has been guided by standard thresholds: values of \( |r| < 0.3 \) have been considered weak, \( 0.3 \leq |r| < 0.7 \) moderate, and \( |r| \geq 0.7 \) strong \cite{Benesty2009}.

Correlation analysis has provided insight into underlying relationships between physical parameters in the dataset, such as velocity dispersion, separation, and stellar mass, without implying causation.

\subsection{Clustering and Nearest Neighbour Search}
Once the set of all predicted WBS was obtained through the ML models, clustering was performed to reduce the sample size for the Nearest Neighbour Search. The K-Means clustering technique was used. K-means clustering partitions a dataset into \( k \) distinct, non-overlapping clusters by minimising the within-cluster sum of squares \cite{Kanungo2002}. The algorithm initializes with \( k \) centroids and iteratively refines their positions by alternating between two steps: assignment of each point to the nearest centroid, and recalculation of centroid positions as the mean of all assigned points. The process converges when centroid positions stabilise or a maximum number of iterations is reached. K-means assumes clusters are spherical and approximately equal in size, which makes it sensitive to outliers and poorly suited for handling non-convex geometries or clusters with variable density. Moreover, the choice of \( k \) must be specified a priori, often guided by heuristics such as the elbow method or silhouette score \cite{Kanungo2002}.

Nearest Neighbour Search (NNS) is a fundamental operation used to identify the closest data point(s) to a given query point in a defined feature space, based on a specific distance metric. It is widely used in applications such as classification, clustering, anomaly detection, recommender systems, and dimensionality reduction.

Given a dataset \( \mathcal{D} = \{x_1, x_2, \ldots, x_n\} \subset \mathbb{R}^d \) and a query point \( q \in \mathbb{R}^d \), the goal of nearest neighbour search (NNS) is to find the point \( x^* \in \mathcal{D} \) minimizing the distance to \( q \), i.e., \( x^* = \arg\min_{x \in \mathcal{D}} \, \text{dist}(q, x) \). Common choices for the distance function include Euclidean distance \( \sqrt{\sum_{i=1}^{d} (q_i - x_i)^2} \), cosine distance \( 1 - \frac{q \cdot x}{\|q\| \, \|x\|} \), or other task-specific metrics \cite{cover1967}. While naive search requires \( \mathcal{O}(n) \) comparisons and becomes expensive for large datasets, efficiency can be improved using methods like KD-Trees (in low dimensions), approximate nearest neighbour (ANN) techniques such as FAISS or HNSW (suitable for high-dimensional data), or locality-sensitive hashing (LSH) to reduce search complexity.

In this study, we apply NNS using 3D Euclidean distance to identify the nearest binary neighbour to each system. This enables us to examine local spatial clustering, detect potential hierarchical or contaminated systems, and provide local density information useful for unsupervised clustering methods like DBSCAN (future work). This spatial NNS is particularly useful in validating the independence of wide binaries in dynamical studies and ensuring that the dataset is not biased by unresolved associations or overlapping systems.

To identify local clustering or spatial associations among binary systems, we employ a nearest neighbour search (NNS) using the three-dimensional (3D) physical distance between binary pairs as the proximity metric. This method is useful for detecting local overdensities, potential contaminants (e.g., hierarchical triples or unbound co-moving pairs), and spatial coherence within a sample.

Given the 3D Cartesian positions \( (x, y, z) \) of stars derived from Gaia parallaxes and sky coordinates, the Euclidean distance between two stars A and B is calculated as:
\begin{equation}
    D_{\mathrm{3D}} = \sqrt{(x_A - x_B)^2 + (y_A - y_B)^2 + (z_A - z_B)^2}.
\end{equation}
For each binary system, we search for its nearest neighbouring binary system in this 3D space. The resulting nearest neighbour distances provide a quantitative measure of local stellar density and can be used to flag potentially non-isolated binaries. In this study, the NNS results are further used in conjunction with clustering algorithms (e.g., DBSCAN) to confirm group memberships and validate the statistical independence of selected binary systems.

It is important to note that the inferred three-dimensional distances, $D_{3D}$, are subject to uncertainties and should not be interpreted as error-free quantities. In the present work, however, these distances are not employed as precise physical measurements but rather to facilitate the ML training process. As such, the method is not strongly dependent on the exact value of $D_{3D}$ as reported in terms of accuracy values in our results, and moderate deviations within the associated confidence intervals do not significantly affect the training outcome. In principle one could add simulated noise in the dataset, however we leave it for future work. 

\section{Wide binary classification problem}
\label{wbcl}
\subsection{The wide binary catalogue} \label{wide-binary catalogue}
In a wide binary, both the components have the same age and composition which make them fit for astronomical analysis \cite{El_Badry_2021}. Moreover, due to their large separations, they help to understand the stellar disk formation in low-density areas \cite{Shaya_2011}. While a wide binary is viewed as two point sources in the sky, chance alignment poses a challenge to the accurate identification of an authentic binary system. For years, the problem of increase in number of chance alignments with the increasing separation has been dealt with through different approaches like the inclusion of proper motion \cite{Chaname_2004,dhital2010} and using parallaxes and radial velocities \cite{Andrews_2017}. 
The subsequent Gaia data releases \cite{Gaia_collab_2018,Gaia_collab_2020,Gaia_collab_2023} has revolutionized the construction of wide binary catalogues by dramatically increasing sample sizes and enabling more precise measurements of parallaxes and proper motions. Based on Gaia DR2 data, several catalogues were formed with different cuts on the separation and parallaxes \cite{elbadry2018,Tian_2020,Hartman_2020}. Similarly, from the Gaia eDR3 dataset, wide binary catalogues have been generated in \cite{El_Badry_2021,Chae_2023}. 
The objective of our work is to predict wide binary pairs from raw Gaia DR3 dataset using a machine learning model trained on the existing catalogue in \cite{El_Badry_2021,el-badry_2021_repo}.

This catalogue makes use of the Gaia eDR3 sources with parallaxes greater than 1 mas, fractional parallax uncertainties less than 20$\%$, absolute parallax uncertainties less than 2 mas, and non-missing G-band magnitudes. The resulting dataset is then subjected to the conditions, following the results of El Badry \cite{El_Badry_2021}:

\begin{itemize}
    \item Projected separation condition. The projected separation should follow:
\begin{equation}{\label{eq:dist}}
    s\leq1pc. \nonumber 
\end{equation}
Equivalently, in terms of the angular separation $\theta$ and the parallax $\tilde{\omega}$:
\begin{equation}
    \theta ~arcsec \leq 206.265\times \tilde{\omega} ~mas
\end{equation}
where $1 ~mas = 10^{-3} arcsec$. 

% The upper limit on the separation is called the Jacobi radius beyond which the Galactic tidal field becomes comparable to the gravitational attraction of the two stars and is given as $r_J\thickapprox 1.35pc \times(\frac{M_{tot}}{M_o})$ (Binney et.al \cite{binney2008}).

    \item Parallax condition due to chance alignment. The condition on the difference between the parallaxes is given by,
    \begin{equation}
   | \tilde{\omega}_1-\tilde{\omega}_2|<b\sqrt{\sigma^2_{\tilde{\omega},1}+\sigma^2_{\tilde{\omega},2}}
    \end{equation}
    where $\sigma_{\bar\omega,i}$ is the parallax uncertainty of the \emph{i}-th component.

In Ref. \cite{El_Badry_2021}, the choice $b=6$ for $\theta<4$ arcsec is adopted to minimize instances of chance alignments.
     
    \item Orbital proper motion. Following Refs. \cite{elbadry2018b,El_Badry_2021} the condition for the difference in proper motion of the two stars to be consistent with a bound Keplerian orbit, amounts to requiring that all candidate binaries have proper motion differences within $3\sigma$ of the maximum velocity difference expected for a system of total mass $5M_\odot$ with circular orbits:
\begin{equation}
        \Delta\mu = [(\mu_{\alpha,1}^*-\mu_{\alpha,2}^*)^2+(\mu_{\delta,1}-\mu_{\delta,2})^2]^{1/2} \leq \Delta{\mu}_{orbit}+3\sigma_{\Delta\mu}.
    \end{equation}
$\Delta{\mu}_{orbit}$ in the above equation is given by,
    \begin{equation}
        \Delta\mu_{orbit}[mas/yr]\leq 0.44(\bar\omega[mas])^{3/2}(\theta[arcsec])^{-1/2};
    \end{equation}
and,
    \begin{equation}
    \begin{split}
        \sigma_{\Delta\mu}=\frac{1}{\Delta\mu}{[(\sigma_{\mu_{\alpha,1}}^*-\sigma_{\mu_{\alpha,2}}^*)\Delta\mu_\alpha^2+(\sigma_{\mu_{\delta,1}}-\sigma_{\mu_{\delta,2}})\Delta\mu_\delta^2]}^{1/2},
    \end{split}
    \end{equation}
where, $\Delta\mu_{\alpha}^2 = (\mu_{\alpha,1}^*-\mu_{\alpha,2}^*)^2$ and $\Delta\mu_{\delta}^2 = (\mu_{\delta,1}-\mu_{\delta,2})^2$.
Subsequent steps include dissolving clusters and cleaning the background to get unbound systems out of moving groups and star clusters, to finally generate the catalogue of labeled wide binaries.
\end{itemize}

\subsection{Methodology}
The objective is to predict whether a particular entry in the raw Gaia dataset is a part of WBS standalone or not (using ML), and further to generate pairs of WBS using Clustering Techniques and 
 Nearest Neighbour Search. 

The catalogue chosen for marking the WBS is a subset of the El Badry catalogue. The code for generating the catalogue can be found in the following Zenodo repository: \href{https://zenodo.org/records/4435257}{Wide-binaries-from-Gaia-eDR3} \cite{el-badry_2021_repo} written for the paper "A Million WBS from Gaia eDR3" by El Badry et al.  \cite{El_Badry_2021}.

\begin{figure}[h!]
\begin{adjustbox}{width=\linewidth,center}
\begin{tikzpicture}[node distance=1.5cm]
% Nodes
\node (start) [startstop] {Obtain Raw Gaia Data};
\node (adql) [process, below of=start] {Obtain the WBS Catalogue from ADQL};
\node (mark) [process, below of=adql] {Mark Target WBS in the Raw Gaia Data};
\node (preprocess) [process, below of=mark] {Preprocess and Filter the Data};
\node (ml) [process, below of=preprocess] {Implement the ML Model};
\node (accuracy) [process, below of=ml] {Perform Accuracy Measurements};
\node (decision) [decision, below of=accuracy, yshift=-1cm] {Is Accuracy Satisfactory?};
\node (tuning) [process, right of=decision, xshift=5cm] {Hyperparameter Tuning};
\node (clustering) [process, below of=decision, yshift=-1cm] {Clustering Techniques};
\node (nns) [process, below of=clustering]{Nearest Neighbour Search};
\node (end) [startstop, below of=nns] {End Process};
% Arrows
\draw [arrow] (start) -- (adql);
\draw [arrow] (adql) -- (mark);
\draw [arrow] (mark) -- (preprocess);
\draw [arrow] (preprocess) -- (ml);
\draw [arrow] (ml) -- (accuracy);
\draw [arrow] (accuracy) -- (decision);
\draw [arrow] (decision) -- (tuning); 
\draw [arrow] (tuning.north) -- ++(0,3.5) -- (ml.east); 
\draw [arrow] (decision.south) -- (clustering.north); 
\draw [arrow] (clustering) -- (nns);
\draw [arrow] (nns) -- (end);
% Labels
\node at (decision.east)[yshift=0.5cm] {No};
\node at (decision.south)[xshift=0.5cm, yshift=-0.25cm] {Yes};
\end{tikzpicture}
\end{adjustbox}
\caption{Methodology for predicting WBS}
\label{fig:wbs-pred}
\end{figure}
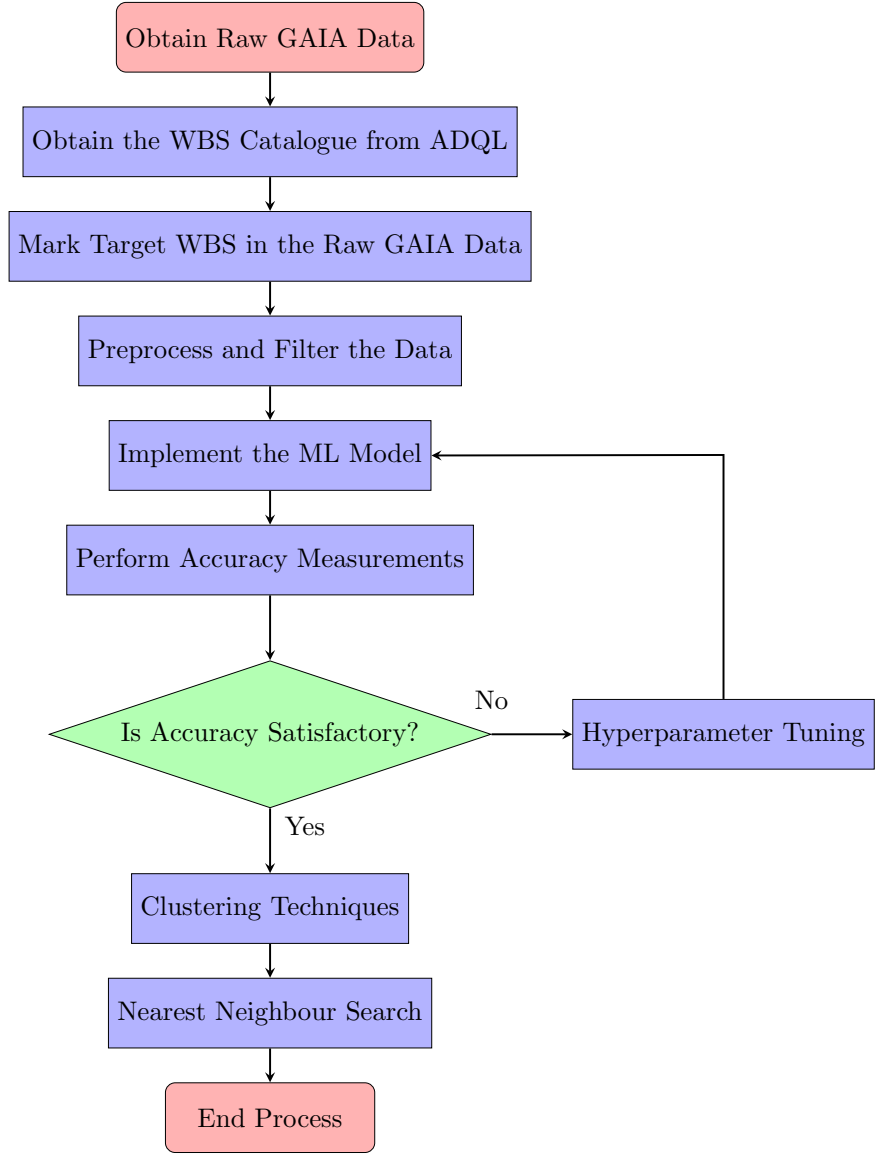

The workflow followed for the problem is outlined in FIG. \ref{fig:wbs-pred}. 

The first step is the process of data extraction from the Gaia data archive. The data is then filtered and made into a catalogue of Wide Binary Systems (WBS) using Astronomical Data Query Language (ADQL). This task is achieved by systematically following the guidelines mentioned in the work done by Banik et al. \cite{Banik_2023}. The next step involves marking the obtained WBCs in the raw dataset. This is done so as to create a label that acts as the target variable of the ML model, and the other features in the dataset are the predictors that are used to train the ML model. The next step is the pre-processing and filtering of data to choose the optimal features for ML classification. The positional information, such as right ascension and declination, is intentionally taken out so as to avoid overfitting. After pre-processing, the task of implementing the ML model and performing accuracy measurements is carried out. According to the accuracy requirements, the hyperparameter values are tuned and the optimal set is selected. Finally, the ML model is used the predict all the WBS in the dataset and then the WBS are paired together using clustering algorithms. 

\FloatBarrier
\FloatBarrier

The following was the flow of the program for the problem:
\begin{enumerate}[label=(\roman*)]
    \item Loading: The first process is loading the raw data and the WBS catalogue. It is worth noting that this WBS catalogue, besides containing the raw features, also contains certain features that are obtained through statistical modelling, for example: 'pm1', 'pm2', 'pmra1', 'pmra2', 'pmdec1', 'pmdec2', 'pairdistance', 'sep\_AU', 'binary\_type'; and many others. 

    \item Labelling: A set called "source\_ids\_set" is constructed from the  'source\_id1' and 'source\_id2' columns of the WBS catalogue. The "source\_ids\_set" is then mapped onto the raw Gaia data to mark the WBS. 

    \item Preprocessing: A check for NULL value containing columns was performed, and those columns were removed, and a filtered dataset was formed. A SMOTE balanced dataset was also generated from this filtered dataset to be used for training the ML models, along with the filtered dataset. 

    Additional steps that can be performed at this stage include: PCA reduction, correlation-based filtering: only including the highly correlated features or setting some correlation cutoff.
    
    \item Implementation: The filtered and the SMOTE-balanced dataset were split into training and test datasets with a 80:20 ratio. The classes are skewed because the data is quite sparse. Therefore, there was a requirement for the SMOTE-balanced dataset to reduce the bias of the ML models during the training process and increase the accuracy and the number of true positives detected. The ML models were tested on the actual dataset to avoid introducing any bias in the system by SMOTE. A variety of ML algorithms were trained (some were trained only on the filtered dataset and not the SMOTE-balanced dataset): Random Forest Classifier (RFC), Logistic Regression (LR), Support Vector Machine with the Radial Basis Function (RBF) kernel (SVM\_RBF), Decision Tree Classifier (DTC), K-Nearest Neighbour Classifier (KNN), Naive Bayes Classifier (NB), Bagging Classifier. 

    \item Evaluation: A threefold evaluation scheme was followed for every ML model: Firstly, the performance metrics were calculated for every model, including accuracy, precision, recall and F1 score. Secondly, the confusion matrices were calculated for every model. And lastly, another table depicting the accuracy of true positives was evaluated that contained the number of true positives, the	true positive rate (\%), the misclassification count and the misclassification rate (\%).
    
    \item Clustering: Once the predictions from the ML model were obtained, the list of objects that were a part of a Wide Binary System was formed. This was the set of WBSs that had to be paired up with their companion stars. To efficiently handle the task of Nearest Neighbour Search and to reduce the computational complexity, clustering techniques were used to divide the data into discrete clusters, and then NNS was performed on each of these clusters. K-Means clustering was performed on the features 'ra' and 'dec' (spatial distance) and 'parallax' (parallax distance). The number of clusters was set to 10.
    
    \item Nearest Neighbour Search: For each of the clusters, NNS was employed to efficiently search for the binary pairs. Given the 3D Cartesian positions \( (x, y, z) \) of stars derived from Gaia parallaxes and sky coordinates, the Euclidean distance between two stars A and B is calculated as:
    \[
    D_{\mathrm{3D}} = \sqrt{(x_A - x_B)^2 + (y_A - y_B)^2 + (z_A - z_B)^2}.
    \]
    For each binary system, the search for its nearest neighbouring binary system in this 3D space was performed. The resulting nearest neighbour distances provide a quantitative measure of local stellar density and can be used to flag potentially non-isolated binaries.
\end{enumerate}
It is important to note that the sole purpose of the nearest neighbour search is to find pairs of stars constituting wide binaries within the list of predicted wide binary systems. In general, for example in tests of gravity, nearest neighbour search should also include nearest neighbour single stars to avoid contamination.    

\subsection{Performance Report and Results}
The SMOTE-balanced-trained ML models are expected to perform better because they reduce the inherent bias of the model due to the sparse dataset. Through TABLE \ref{tab:wbs-perform-rfc}, it is clear that the SMOTE-balanced models perform much better than the base models. 

More so, it is clear from TABLE \ref{tab:wbs-class-rfc} that there is an extremely high rate of misclassification in the base model, which is extremely reduced in the SMOTE-balanced dataset.

\setlength{\tabcolsep}{2pt}
\renewcommand{\arraystretch}{1.5}
\begin{table}[h]
\centering
\caption{Performance metrics of the RFC algorithm on the raw-filtered dataset and SMOTE-balanced dataset for WBS detection}
\label{tab:wbs-perform-rfc}
\begin{tabular}{|c|c|c|c|c|}
\hline
\textbf{Algorithms} & \textbf{Precision} & \textbf{Recall} & \textbf{F1 score} & \textbf{Accuracy} \\
\hline
RFC & 0.375000 & 0.008234 & 0.016115 & 0.98901 \\
\hline
RFC (SMOTE) & ~0.917273~ & ~0.923147~ & ~0.920201~ & ~0.99825~ \\
\hline
\end{tabular}
\end{table}

\setlength{\tabcolsep}{1pt} % Column padding
\renewcommand{\arraystretch}{1.5}
\begin{table}[h]
\centering
\caption{Classification Analysis of the RFC Algorithm on the raw-filtered dataset and SMOTE-balanced dataset for WBS detection}
\label{tab:wbs-class-rfc}
\begin{tabular}{|c|c|c|p{1.5cm}|p{2cm}|}
\hline
\textbf{Algorithms} & \textbf{TP} & \textbf{TP rate (\%)} & \textbf{Misclass- ifications} & \textbf{Misclassifica- tion rate(\%)} \\
\hline
RFC & 9 & 0.823422 & 1099 & 100.548948 \\
\hline
RFC (SMOTE) & 1009 & 92.314730 & 175 & 16.010979 \\
\hline
\end{tabular}%
\end{table}
\FloatBarrier

\begin{figure}[h!]
    \centering
    \subfloat[Raw predictions\label{fig:wbs-cm-raw-rfc}]{
        \includegraphics[width=0.4\linewidth]{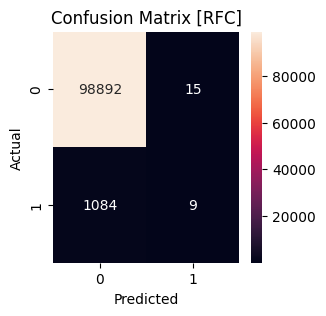}
    }
    \hfill
    \subfloat[SMOTE-balanced predictions\label{fig:wbs-cm-smote-rfc}]{
        \includegraphics[width=0.45\linewidth]{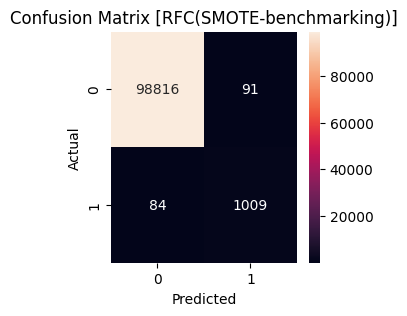}
    }
    \caption{Confusion matrices for the raw-filtered dataset predictions and the SMOTE-balanced dataset predictions}
\end{figure}

The confusion matrices also showcase the low detection rate of the base models, which is improved by the SMOTE-balanced models as observed in the FIG. \ref{fig:wbs-cm-raw-rfc} and FIG. \ref{fig:wbs-cm-smote-rfc}. A detailed plot for all the tested algorithms is given in the appendix \ref{app:WBS}. 
\FloatBarrier

% \begin{figure}[h!]
%     \centering
%     \begin{subfigure}{0.3\linewidth}
%         \centering
%         \includegraphics[width=\linewidth]{WBS_CM_raw_RFC.png}
%         \caption{Raw predictions}
%         \label{fig:wbs-cm-raw-rfc}
%     \end{subfigure}
%     \hfill
%     \begin{subfigure}{0.4\linewidth}
%         \centering
%         \includegraphics[width=\linewidth]{WBS_CM_SMOTE_RFC.png}
%         \caption{SMOTE-balanced predictions}
%         \label{fig:wbs-cm-smote-rfc}
%     \end{subfigure}
%     \caption{Confusion matrices for the raw-filtered dataset predictions and the SMOTE-balanced dataset predictions}
% \end{figure}
\FloatBarrier

The clustering was performed with the total number of clusters set to ten. The nearest neighbour search was performed on each of the ten clusters to find the binary pairs. 
\begin{figure}[!ht]
    \centering
    \includegraphics[width=0.9\linewidth]{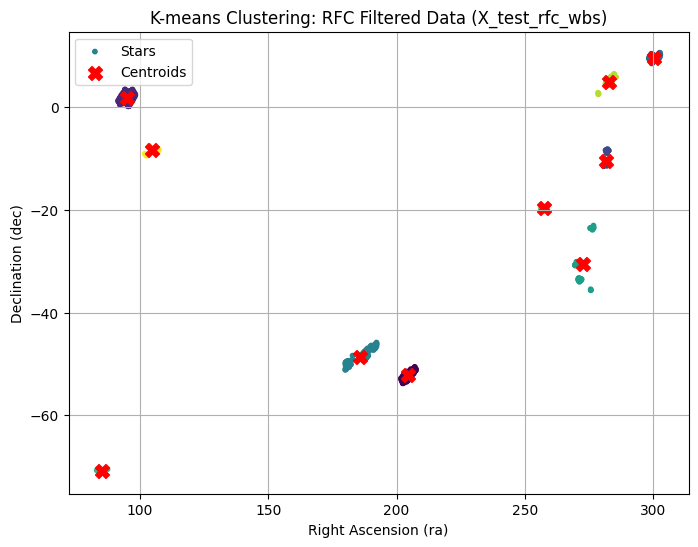}
    \caption{The distribution of the clusters}
    \label{fig:clustering}
\end{figure}
\begin{figure}
    \centering
    \includegraphics[width=0.9\linewidth]{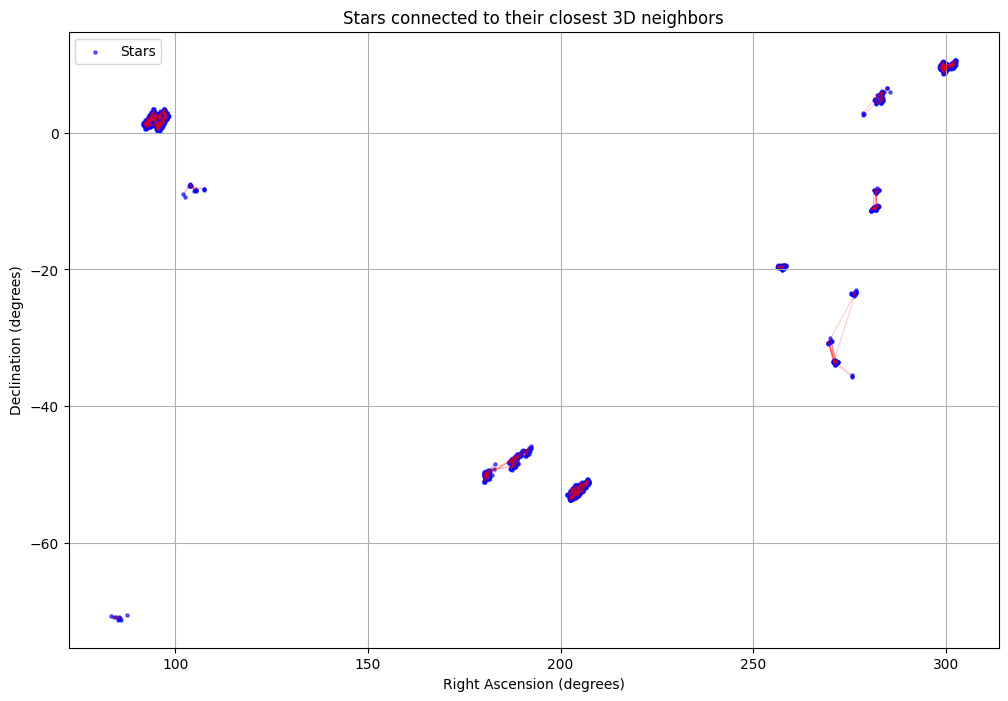}
    \caption{WBS connected to their respective pairs}
    \label{fig:nns}
\end{figure}
\FloatBarrier

\section{Conclusions}
\label{conclusion}
We have applied machine learning (ML) techniques to the problem of the detection of wide binaries. Wide binary stars—stellar pairs separated by hundreds to thousands of astronomical units—are crucial astrophysical laboratories for testing gravitational theories, including potential deviations from Newtonian dynamics at large separations. 

While conventional statistical techniques for detecting wide binaries are computationally expensive, relying on Monte-Carlo simulations and complex probabilistic analyses to rule out chance alignments, a machine learning based approach offers a scalable alternative where techniques such as clustering algorithms and nearest neighbour search are used to efficiently predict binaries from noisy background populations. In our implementation, the catalogue generated by El-Badry \etal \cite{el-badry_2021_repo} is the primary training dataset for the analysis of Sec. \ref{wbcl}. However, the trained model can be used to predict pairs of a wide binary system directly from the raw Gaia DR3 data.  

The publicly available tool (hosted at \href{https://github.com/DespCAP/G-ML}{https://github.com/DespCAP/G-ML}) developed as part of this work can be used to generate a catalogue of Wide Binary Stars quickly and fairly accurately from the raw Gaia source data. The type of ML model to be used and its hyperparameters, the kind of preprocessing techniques to be used, and clustering criteria are all tunable by the user. Our code also allows for tuning the marking process in the training phase, which enables the user to substitute a better alternative which can more accurately determine the wide binary systems. There is also a provision for importing the pretrained model parameters so that the training phase is skipped, and the user can directly use the tool to predict WBS and make catalogues.

In summary, the program provides a list of WBS based on Raw Gaia data, without mapping them to their respective pairs. The trained model(s) can be directly used for predictions, which skips the training phase. The provision of a transfer-learning compatible solution ensures the ML training can be outsourced and used by a wider class of people. This automates the tedious analytical and statistical process of finding WBS systems.

As part of an ongoing work, we plan to extend the scope of application of machine learning in the context of wide binaries to identify potential anomalous wide binaries by casting it as a supervised anomaly detection problem. This task would be a step towards using ML models to learn the characteristics of systematic deviations from newtonian gravity in a population sample of wide binaries as reported in recent literature \cite{Hernandez2021,hernandez2024,Hernandez_2012,Hernandez2023,Hernandez2023b,chae2023robustevidencebreakdownstandard,chae2024,Chae_2022,Chae_2023}. We foresee several interesting directions that can be taken up as future problems, such as integrating and merging the repository on WBS predictions and anomaly detection on the raw Gaia data, expanding the models to predict more exotic and general gravitational phenomena or building an ML-based stellar object identifier for the Gaia data.

%\centering{\textbf{References}}
\bibliographystyle{apsrev4-1}
\bibliography{ref}

\appendix

\section{Predicting WBS using ML}
\label{app:WBS}
\subsection{Data characteristics}
The class distribution for the target column in the filtered dataset was:
    \begin{verbatim}
    y.value_counts()
        0    494664
        1      5336
    \end{verbatim}
The class distribution for the target column in the training and test datasets of the filtered dataset were:
    \begin{verbatim}
    y_train.value_counts()
        0    395757
        1      4243

    y_test.value_counts()
        0    98907
        1     1093
    \end{verbatim}
The class distribution for the target column in the training dataset for the SMOTE-balanced dataset was:
    \begin{verbatim}
    re_y_train.value_counts()
        0    396153
        1    395309
    \end{verbatim}

\subsection{Detailed plots and tables}
\setlength{\tabcolsep}{8pt}
\renewcommand{\arraystretch}{1.3}
\begin{table}[ht]
\centering
\caption{Performance comparison of different ML algorithms on the raw-filtered dataset for WBS detection}
\label{tab:wbs-perform-raw}
\begin{tabular}{|c|c|c|c|c|}
\hline
\textbf{Algorithms} & \textbf{Precision} & \textbf{Recall} & \textbf{F1 score} & \textbf{Accuracy} \\
\hline
RFC & 0.375000 & 0.008234 & 0.016115 & 0.98901 \\
\hline
LR & 0.000000 & 0.000000 & 0.000000 & 0.98907 \\
\hline
SVM (RBF) & 0.000000 & 0.000000 & 0.000000 & 0.98907 \\
\hline
DTC & 0.116667 & 0.134492 & 0.124947 & 0.97941 \\
\hline
AdaBoost & 0.400000 & 0.001830 & 0.003643 & 0.98906 \\
\hline
KNN & 0.000000 & 0.000000 & 0.000000 & 0.98907 \\
\hline
NB & 0.024346 & 0.086002 & 0.037949 & 0.95234 \\
\hline
Bagging & 0.222222 & 0.018298 & 0.033812 & 0.98857 \\
\hline
\end{tabular}
\end{table}

\setlength{\tabcolsep}{6pt}
\renewcommand{\arraystretch}{1.3}
\begin{table}[ht]
\centering
\caption{Performance Comparison of Machine Learning Algorithms with SMOTE balanced dataset for WBS detection}
\label{tab:wbs-perform-smote}
\begin{tabular}{|c|c|c|c|c|}
\hline
\textbf{Algorithms} & \textbf{~Precision~} & \textbf{~Recall~} & \textbf{~F1 score~} & \textbf{~Accuracy~} \\
\hline
RFC(SMOTE) & ~0.917273~ & ~0.923147~ & ~0.920201~ & ~0.99825~ \\
\hline
LR(SMOTE) & ~0.024428~ & ~0.086002~ & ~0.038049~ & ~0.95247~ \\
\hline
DTC(SMOTE) & ~0.668024~ & ~0.900274~ & ~0.766952~ & ~0.99402~ \\
\hline
AdaBoost(SMOTE) & ~0.061118~ & ~0.493138~ & ~0.108757~ & ~0.91166~ \\
\hline
KNN(SMOTE) & ~0.039785~ & ~0.867338~ & ~0.076080~ & ~0.76975~ \\
\hline
NB(SMOTE) & ~0.024093~ & ~0.085087~ & ~0.037553~ & ~0.95233~ \\
\hline
Bagging(SMOTE) & ~0.890291~ & ~0.838975~ & ~0.863872~ & ~0.99711~ \\
\hline
\end{tabular}
\end{table}

\setlength{\tabcolsep}{7pt} % Column padding
\renewcommand{\arraystretch}{1.2}
\begin{table}[ht]
\centering
\caption{Classification Analysis of ML Algorithms on the raw-filtered dataset for WBS detection}
\label{tab:wbs-class-raw}
\resizebox{\textwidth}{!}{%
\begin{tabular}{|c|c|c|c|c|}
\hline
\textbf{Algorithms} & \textbf{TP} & \textbf{TP rate (\%)} & \textbf{Misclassifications} & \textbf{Misclassification rate (\%)} \\
\hline
RFC & 9 & 0.823422 & 1099 & 100.548948 \\
\hline
LR & 0 & 0.000000 & 1093 & 100.000000 \\
\hline
SVM (RBF) & 0 & 0.000000 & 1093 & 100.000000 \\
\hline
DTC & 147 & 13.449222 & 2059 & 188.380604 \\
\hline
AdaBoost & 2 & 0.182983 & 1094 & 100.091491 \\
\hline
KNN & 0 & 0.000000 & 1093 & 100.000000 \\
\hline
NB & 94 & 8.600183 & 4766 & 436.047575 \\
\hline
Bagging & 20 & 1.829826 & 1143 & 104.574565 \\
\hline
\end{tabular}%
}
\end{table}

\setlength{\tabcolsep}{7pt}
\renewcommand{\arraystretch}{1.2}
\begin{table}[!ht]
\centering
\caption{Classification Analysis of ML Algorithms with SMOTE balanced dataset for WBS detection}
\label{tab:wbs-class-smote}
\resizebox{\textwidth}{!}{%
\begin{tabular}{|c|c|c|c|c|}
\hline
\textbf{Algorithms} & \textbf{TP} & \textbf{TP rate (\%)} & \textbf{Misclassifications} & \textbf{Misclassification rate (\%)} \\
\hline
RFC(SMOTE) & 1009 & 92.314730 & 175 & 16.010979 \\
\hline
LR(SMOTE) & 94 & 8.600183 & 4753 & 434.858188 \\
\hline
DTC(SMOTE) & 984 & 90.027447 & 598 & 54.711802 \\
\hline
AdaBoost(SMOTE) & 539 & 49.313815 & 8834 & 808.234218 \\
\hline
KNN(SMOTE) & 948 & 86.733760 & 23025 & 2106.587374 \\
\hline
NB(SMOTE) & 93 & 8.508692 & 4767 & 436.139067 \\
\hline
Bagging(SMOTE) & 917 & 83.897530 & 289 & 26.440988 \\
\hline
\end{tabular}%
}
\end{table}
\FloatBarrier

\FloatBarrier
\begin{figure}[h!]
    \centering
    \includegraphics[width=0.6\linewidth]{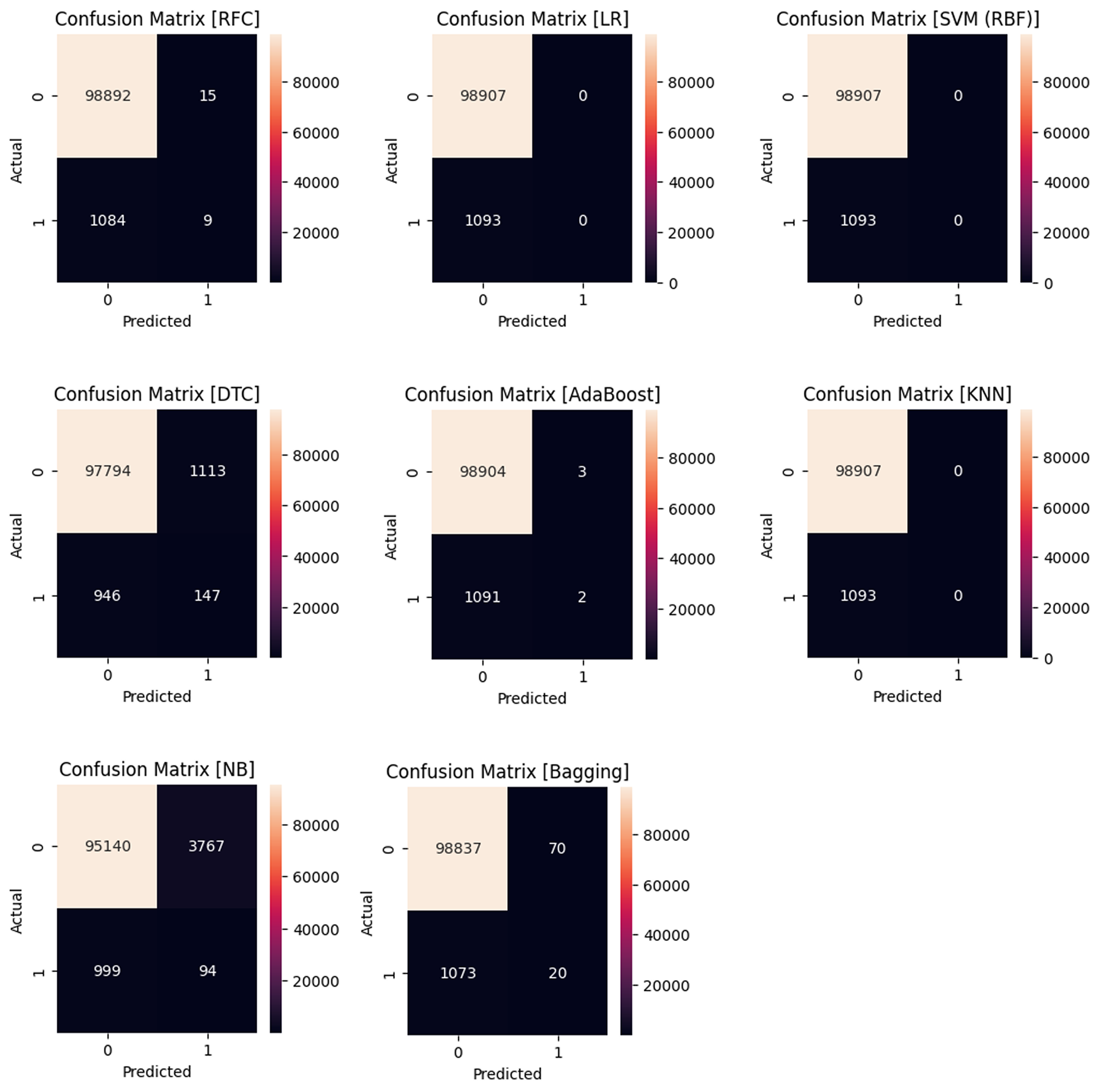}
    \caption{CM for the raw-filtered dataset predictions}
    \label{fig:wbs-cm-raw}
\end{figure}

\begin{figure}[h!]
    \centering
    \includegraphics[width=0.7\linewidth]{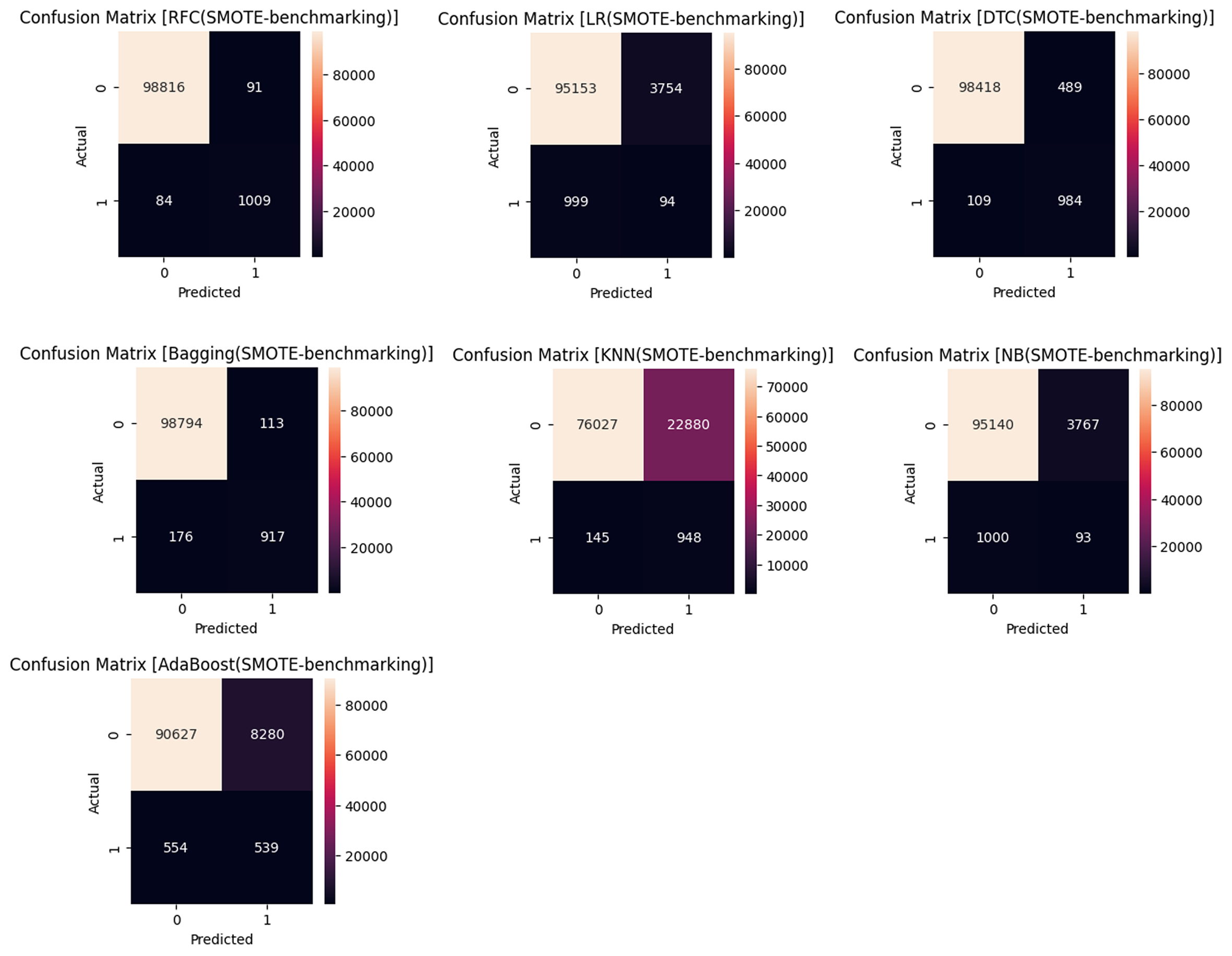}
    \caption{CM for the SMOTE-balanced dataset predictions}
    \label{fig:wbs-cm-smote}
\end{figure}
\FloatBarrier

\end{document}